\def \bc {\begin{center}}
\def \ec {\end{center}}

\def \bfr {\begin{flushright}}
\def \efr {\end{flushright}}

\def \v {\vskip}
\def \ba {\begin{array}}
\def \ea {\end{array}}

\def \bea {\begin{eqnarray}}
\def \eea {\end{eqnarray}}

\def \be {\begin{equation}}
\def \ee {\end{equation}}
\def \d {\hbox{d}\,}%% The exterior derivative operator 
\def \square {\hbox{$\sqcup\!\!\!\!\sqcap$}} %% 

\def \e {\hbox{e}}

\documentstyle[12pt,a4]{article}
\begin{document}
\title{New Symmetries in Two-Dimensional Dilaton Gravity}

\begin{center}

{\bf{\LARGE New Symmetries in Two-Dimensional Dilaton Gravity
\footnote{Work partially supported by the 
{\it Comisi\'on Interministerial de Ciencia y Tecnolog\'{\i}a}\/ 
and {\it DGICYT}.}}
}

\bigskip 

 J.~Cruz$^1$\footnote{\sc cruz@lie.uv.es},
 J.~Navarro-Salas$^1$\footnote{\sc jnavarro@lie.uv.es},
 M.~Navarro$^{2,3}$\footnote{\sc mnavarro@ugr.es} 
 and
 C.~F.~Talavera$^{1,4}$\footnote{\sc talavera@lie.uv.es}

\end{center}

\bigskip% 

\footnotesize
\begin{enumerate}	
\item Departamento de F\'{\i}sica Te\'orica and 
	IFIC, Centro Mixto Universidad de Valencia-CSIC.
	Facultad de F\'{\i}sica, Universidad de Valencia,	
        Burjassot-46100, Valencia, Spain. 
\item Instituto Carlos I de F\'\i sica Te\'orica y Computacional,
        Facultad  de  Ciencias, Universidad de Granada. 
        Campus de Fuentenueva, 18002, Granada, Spain. 
\item Instituto de Matem\'aticas y F\'\i sica Fundamental, 
        CSIC. Serrano 113-123, 28006 Madrid, Spain.
\item Departamento de Matem\'atica Aplicada.      
	E.T.S.I.I. Universidad Polit\'ecnica de Valencia.      
	Camino de Vera, 14. 46100-Valencia, Spain.                  
\end{enumerate}
\normalsize 

\bigskip%\vskip2mm 
%\centerline{\today}
\bigskip%\vskip2mm

\begin{center}
			{\bf Abstract}
\end{center}

\footnotesize 

 We present three types of non-conformal symmetries for a wide class of 2D 
 dilaton-gravity models. 
 For the particular CGHS, or string-inspired model 
	a linear combination of these symmetries is conformal
 and turns out to be the well-known symmetry which allows to construct the 
 exactly solvable semiclassical RST and BPP models. 
	We show that one of these non-conformal symmetries
 can be converted into a conformal one by means of a suitable field 
 redefinition involving the metric and the derivatives of the dilaton. As a 
 consequence of this, and by defining a Polyakov-type term in terms of
an invariant metric under the symmetry, we are able to provide, 
for a generic 2D dilaton
gravity model, associated semiclassical models which are symmetry invariant. 

\normalsize 
\vskip5mm

\centerline{\it Talk given at} 
\centerline{The XXI International Colloquium on Group Theoretical Methods in
Physics} 
\v3mm
\centerline{15--20 July 1996, Goslar, Germany}
\v6mm
\centerline{(To appear in the Proceedings)}
\v2mm
\newpage
\section{Two-dimensional dilaton gravity models}

Nowadays one of the main objectives of Theoretical Physics is to produce a  
quantum theory of gravity. 
The four-dimensional Einstein-Hilbert gravity
theory, with action 

\be S_{\hbox{HE}}= \frac1{16\pi}\int\d^4 x 
\sqrt{-^{(4)}g}{}^{(4)}R + S_{\hbox{M}}\label{EH}\ee  
is, however, very complex to handle. 
Toy models, i.e., models which  
share with Einstein gravity its most 
relevant features but are far simpler to deal with  
should, therefore, play an important role here. 2D dilaton models
(2DDM) are two-dimensional models of gravity which are 
general covariant and involve a dilaton field. These models have  
classical solutions describing the formation of
two-dimensional black holes. Therefore, 2DDMs can be applied  
to study the dynamics of black holes in a  
simplified setting. 

2DDMs have appeared in diverse areas of 
theoretical physics, string theory in particular \cite{Witten}, 
but they can simply be regarded as models  
of gravity in which all but two dimensions 
have been frozen out. Consider for instance four-dimensional
spherically-symmetric Einstein gravity (SSG). 
The metric can be parametrized as 
\be {}^{(4)}\hbox{d} s^2 = {}^{(2)}g_{\mu\nu}\d x^\mu \d x^\nu + 
\frac{\e^{-2\phi}}{\lambda^2}\d \Omega^2\label{ss}\ee  
where ${}^{(2)}g_{\mu\nu}$ and $\phi$ are defined on a  
two-dimensional manifold which can be 
co-ordinatized by $t$ and $r$. If we
place the metric (\ref{ss}) into the Einstein-Hilbert action
(\ref{EH}) we get $\left({}^{(2)}g_{\mu\nu}\rightarrow g_{\mu\nu}\right)$ 

\be S_{\hbox{SSG}}=\frac1{2\pi}\int \d^2x\sqrt{-g}\e^{-2\phi}
\left(R(g) +2g^{\mu\nu}\nabla_\mu\phi\nabla_\nu\phi  
+2\lambda^2\e^{2\phi}\right) 
+ S_{\hbox{M}}\label{ssg}\ee 

2DDMs generalize the SSG model and their general  
action can be written in the form   

\be S_{\hbox{GDG}}\left(g,\phi\right) = {1\over2\pi} \int d^2x\sqrt{-g}
\left[ D(\phi) R + {1\over2} \left(\nabla\phi\right)^2 + 
F(\phi) \right] +  S_{\hbox{M}}\label{GDG}\ee 
where $D$ and $F$ are arbitrary functions. 

A result which is particularly useful 
is that after suitable redefinitions of the 
two-dimensional metric $g_{\mu\nu}$ and the dilaton field $\phi$  
every action can be brought to the form 
\cite{Banks,Gegenberg} 

\be 
\tilde{S}_{\hbox{GDG}}\left(\tilde{g},\tilde{\phi}\right) = 
{1\over2\pi} \int d^2x\sqrt{-\tilde{g}}
\left(\tilde{R}\tilde{\phi} + \tilde{V} (\tilde{\phi})\right) 
+ S_{\hbox{M}}\label{GDG2} \ee 

\newpage
\section{The CGHS model, Hawking radiation 
and the BPP and RST models}

The CGHS or string-inspired model 
of two-dimensional dilaton gravity \cite{Witten,CGHS}

\be S_{\hbox{CGHS}} = {1\over2\pi} \int d^2 x \sqrt{-g}
	\left[ e^{-2\phi} (R + 4 (\nabla\phi)^2 + 4\lambda^2 ) 
-{1\over2} \sum_{i=1}^N (\nabla f_i)^2 \right]\label{CGHS}\ee 
has attracted particular attention because it posseses an extra 
conformal symmetry 

\be\delta \phi =\epsilon \e^{2\phi},\quad 
\delta g_{\mu\nu}=2\epsilon\e^{2\phi}g_{\mu\nu}\label{symmetry}\ee 
which makes the model solvable (see 
\cite{Kazama}). 

In two dimensions Hawking radiation, which is a result 
of the trace anomaly 

\[g_{\mu\nu}<T^{\mu\nu}_{\hbox{M}}>=\frac{N}{24}R,\]   
and its back-reaction on the black-hole geometry 
can be taken into account at a semiclassical level 
by adding to the classical action
$S_{\hbox{CGHS}}$ a Polyakov-Liouville  term

\be 
S_{\hbox{P}} = - {N\over96\pi} \int d^2x\sqrt{-g} R\square^{-1}R 
\ee 
However, since this term is not invariant under the symmetry
(\ref{symmetry}), the semiclassical action 
$S_{\hbox{CGHS}}+S_{\hbox{P}}$ is neither invariant nor solvable.
 
This problem can be solved as follows. 
Suppose we are able to find a metric
$\bar{g}_{\mu\nu}=\bar{g}_{\mu\nu}\left(g_{\mu\nu},\phi\right)$ and a
dilaton field $\bar\phi$  such that 
$\delta\>\bar{g}_{\mu\nu}=0$ . 
If now we express the action $S_{\hbox{CGHS}}$ in terms of the barred 
variables and add the Polyakov-Liouville term also written 
in terms of the barred metric, 
the resultant semiclassical action will clearly be
symmetry invariant. Going back to the unbarred quantities by 
undoing the change of variables we end
up with a semiclassical action
$S_{\hbox{CGHS}}+S_{\hbox{P}} +S'_{\hbox{P}}$ which is symmetry 
invariant, being $S'_{\hbox{P}}$ a symmetry-restoring counterterm. 

For the CGHS model, it is easy to see that the tilded functions 
\[
\tilde{g}_{\mu\nu} = g_{\mu\nu} e^{-2\phi} \>, \quad 
\tilde{\phi} = e^{-2\phi} \>\]
which bring the action (\ref{CGHS}) into the form (\ref{GDG2}) with 
$\tilde{V}=4\lambda^2$   
can be used as barred variables. 
By using these barred quantities, the above-described 
procedure gives rise to 
the following semiclassical symmetry-invariant action 
\be{S}_{\hbox{Sem}}(g,\phi)= {S}_{\hbox{CGHS}}(g,\phi) + 
S_{\hbox{P}}(g)+ S_{\hbox{BPP}}(g,\phi)\label{BPP}\ee 
with 
\be S_{\hbox{BPP}}= {N\over24\pi} \int d^2 x \sqrt{-g} 
\left( (\nabla\phi)^2 - \phi R \right)\ee 
which is the BPP model \cite{BPP}. A new change of variables 

\be e^{-2\phi} = e^{-2\phi'} +\frac{N}{24}\phi',\qquad 
g_{\mu\nu}=\frac{g'_{\mu\nu}}{1 +\frac{N}{24}\phi'e^{2\phi'}}
\nonumber\ee 
yields the RST model \cite{RST}.

For completeness we would like to mention that the BPP 
action (\ref{BPP}) can also be obtained by (formally) 
promediating the action 
${S}_{\hbox{CGHS}}(g,\phi) + S_{\hbox{P}}(g)$ over the 
one-parameter group generated by the symmetry (\ref{symmetry}).
 
\section{New symmetries in generic 2D dilaton gravity}

Now we shall see that what have been 
done in the previous section with the CGHS model can also
be done  -- with suitable modifications -- with a
generic 2D dilaton model of gravity. 

Leaving aside the matter term which will be introduced later,  
 the general action for 2DDMs  
(\ref{GDG2}) is invariant under the following symmetries 
(the tildes will be dropped from now on) 

\bea 
\delta_E \phi=0 
&, & 
\delta_E g_{\mu\nu}= 
g_{\mu\nu}a_\sigma\nabla^\sigma\phi-\frac12 
\left( a_\mu \nabla_\nu\phi + a_\nu \nabla_\mu \phi \right)
\> \\
\delta_1\phi=0 
&,& 
\delta_1 g_{\mu\nu}=
\epsilon_1\left({g_{\mu\nu}\over\left(\nabla\phi\right)^2}
-2{\nabla_{\mu}\phi\nabla_{\nu}\phi\over\left(\nabla\phi\right)^4}\right)
\> \nonumber\\
\delta_2\phi = \epsilon_2
&,&
\delta_2 g_{\mu\nu}=\epsilon_2 V\left({g_{\mu\nu}\over\left(\nabla\phi
\right)^2}-2{\nabla_{\mu}\phi\nabla_{\nu}\phi
	\over\left(\nabla\phi\right)^4}\right) 
\>\label{gensym}\\
\delta_3\phi=0 &,&
\delta_3 g_{\mu\nu}= -\frac{\epsilon_3}2\left[g_{\mu\nu}+ 
J\left({g_{\mu\nu}\over\left(\nabla\phi\right)^2}
-2{\nabla_{\mu}\phi
\nabla_{\nu}\phi\over\left(\nabla\phi\right)^4}\right)\right]
\> \nonumber
\eea 
The Noether currents are, respectively 

\be
J^{\mu\nu}= g^{\mu\nu}\frac12\left((\nabla\phi)^2 -J(\phi)\right)
\equiv g^{\mu\nu}E 
\qquad \left(J^{\prime}\left(\phi\right)=V(\phi)\right)\ee 

\be j_{1}^{\mu}={\nabla^{\mu}\phi\over\left(\nabla\phi\right)^2}
\>,\quad
 j_2^{\mu}=j_R^{\mu}+V{\nabla^{\mu}
\phi\over\left(\nabla\phi\right)^2}\>,\quad
j_3^{\mu }=Ej_1^{\mu}\label{currents}
\ee
In the limiting case $V=4\lambda^2$ the symmetry $\delta =  
\delta_2 -4\lambda^2\delta_1$ is the conformal symmetry
(\ref{symmetry}) of the string inspired  model.  It is apparent, 
therefore, that we have generalized the symmetry of the CGHS model 
to an arbitrary 2D dilaton gravity
model  -- albeit the symmetry is no longer 
conformal\footnote{The most general model which allows a
conformal symmetry is the exponential model with 
$V\propto e^{\beta\phi}$ \cite{symbh}}. 

Now we can apply the above-described procedure to produce a
semiclassical action which is invariant under $\delta$. It reads 

\be
S = S_{\hbox{GDG}}\left[g(\bar{g},\bar{\phi}),\phi(\bar{\phi})\right]
- {1\over2} \sum_{i=1}^N \int d^2 x \sqrt{-\bar{g}} \bar{g}^{\mu\nu}
\partial_\mu f_i \partial_\nu f_i 
+ S_P (\bar{g})
\> \label{invact}
\ee
where $\bar{\phi}\equiv\phi$ and 
 
\be \bar g_{\mu\nu} = 
\frac{2E_\lambda}{\left(\nabla\phi\right)^2}g_{\mu\nu}+
\left(\frac1{2E_\lambda}-{\frac{2E_\lambda} 
{\left(\nabla\phi\right)^4}}\right)
\nabla_{\mu}\phi\nabla_{\nu}\phi\label{Emetric}\ee
with  
\[E_\lambda = \frac12\left((\nabla\phi)^2 -J(\phi)  
+4\lambda^2\phi\right)\equiv E+2\lambda^2\phi\]
Eq. (\ref{Emetric}) is the simplest  metric which  
is invariant under $\delta =\delta_2-4\lambda^2\delta_1$ and such that 
$\bar g_{\mu\nu}\equiv g_{\mu\nu}$ when $V=4\lambda^2$, and  
det$\ g_{\mu\nu}$ $=$ det$\ \bar g_{\mu\nu}\ $. 
Therefore the action (\ref{invact}) 
will reduce to the BPP model
(\ref{BPP}) for $V = 4\lambda^2$.
Notice, however, that up to the CGHS and exponential models 
the action (\ref{invact}) contains second derivatives 
and that the counterterm added to the
Polyakov-Liouville term  to preserve the symmetry 
is necessarily non-local. 

In conclusion, we have provided a semiclassical 
action for generic 2DDM which is invariant under the transformation
\be
\delta \bar\phi = \epsilon \>, \qquad
\delta \bar{g}_{\mu\nu} = 0 \>
\ee
This is the standard expression for the symmetry of the CGHS model 
which allows to reduce the associated sigma model to a (solvable) 
Liouville-type theory.
We expect that a generalization of the approach of Ref.~\cite{Kazama} 
to a
second order non-linear sigma model could imply solvability of the
present theory, in terms of the invariant metric.

\section*{Acknowledgments}
M.~N. is grateful to the Spanish MEC, CSIC and IMAFF (Madrid)
for a research contract.

\end{document}